\documentclass{blois}

\def\be{\begin{equation}}
\def\ee{\end{equation}}
\def\bea{\begin{eqnarray}}
\def\eea{\end{eqnarray}}

\makeatletter
\def\@xfootnote[#1]{%
  \protected@xdef\@thefnmark{#1}%
  \@footnotemark\@footnotetext}
\makeatother

\newcommand{\Photo}{}

\begin{document}
\vspace*{4cm}
\title{LIGHT STOP DECAYS}

\author{ RAMONA GR\"{O}BER$^1$
\footnote[*]{Talk given at the 27th Rencontres de Blois on Particle Physics and Cosmology, May 31 - June 05, 2015.}
}

\address{$^1$INFN, Sezione di Roma Tre, Via della Vasca Navale 84,\\
I-00146 Roma, Italy}

\maketitle\abstracts{
If the stop is the next-to-lightest supersymmetric particle (NLSP) and the mass difference to the neutralino is smaller than the top mass, it can decay via flavour-violating decay modes to $c\tilde{\chi}^0_1/u\tilde{\chi}^0_1$ or a four-body decay to $b\tilde{\chi}_1^0f\bar{f}'$, which above the $W$ boson threshold corresponds to the decay to $b\tilde{\chi}_1^0W$. Improving on existing calculations for these decay modes, we analyse the branching ratios (BRs) for the respective decays and  show that they can significantly deviate from one. }

\section{Introduction}
While the limits for squarks of the first and second generations are pushed already above the 1 TeV range, lighter stops are not yet completely excluded. In particular, in the mass range where the mass difference to the neutralino is below the top mass, experimental searches are very challenging.  In this kinematic region, assuming the neutralino  $\tilde{\chi}_1^0$ to be the lightest supersymmetric particle (LSP), the stop can decay via flavour violating decays to $c\tilde{\chi}_1^0/u\tilde{\chi}_1^0$ or via a four-body decay to $b\tilde{\chi}_1^0f\bar{f}'$, or above the $W$ boson threshold to $b\tilde{\chi}_1^0W$. For the latter decays ATLAS and CMS provide limits in searches for one or two leptons and missing energy \cite{Aad:2014qaa}  \cite{Aad:2014kra}, for the four-body decays in searches with one isolated lepton, jets and missing energy \cite{Aad:2014kra} or in monojet searches \cite{ATLAScharm} and for the decays to $c\tilde{\chi}_1^0$ in monojet searches and charm-tagged searches \cite{ATLAScharm}.
\par
 Branching ratios of the stop in the respective decay channels can often differ from one. Only since recently the experimental collaborations account for that \cite{Aad:2015pfx}. After discussing flavour violation (FV) in the minimal supersymmetric extension of the Standard Model (MSSM) in sec.~\ref{flavour}, the computation of the decay widths and BRs of the aforementioned decays are reviewed in sec.~\ref{decays}. In sec.~\ref{numana} numerical results are shown and compared to experimental exclusion bounds, before concluding in sec.~\ref{conclusion}.

\section{Flavour violation in the MSSM \label{flavour}}
The MSSM provides many new sources of FV. However, these sources are strongly restricted by flavour experiments. This so-called "new physics flavour puzzle" can e.g.~be solved by  Minimal Flavour Violation (MFV) \cite{Chivukula:1987fw}. In MFV the Lagrangian is constructed such that it is formally invariant under a flavour $SU(3)_{Q_L}\times SU(3)_{u_R}\times SU(3)_{d_R}$ symmetry, by promoting the Yukawa couplings to spurions. Hence the only source of FV is restricted to the Yukawa couplings. Even reduced flavour symmetries can still be in accordance with flavour experiments \cite{Barbieri:2012bh}. 
In particular, to obtain one lighter stop, we will allow for a smaller soft-SUSY breaking mass $m_{\tilde{t}_R}$. This will hence reduce the flavour symmetry of the right-handed up sector to $SU(2)_{u_R}$. Furthermore, we distinguish between two cases, namely whether the $SU(3)_{Q_L}$ flavour symmetry is reduced to a $SU(2)_{Q_L}$ or not (dubbed with $U(2)$ or $U(3)$, respectively). Note that even if at one scale the mixing matrix of the squarks is chosen flavour diagonal, this does not hold true at any other scale. Renormalisation group running induces FV at any other scale.
For further reference, we assume that the lightest mass eigenstate $\tilde{u}_1$ is mainly stop-like and hence call it stop, but having in mind that it is an admixture of all flavour eigenstates.  
\section{Decays of a light stop \label{decays}}
 \subsection{Flavour-violating decays}
\begin{figure}
 \includegraphics[width=7.7cm]{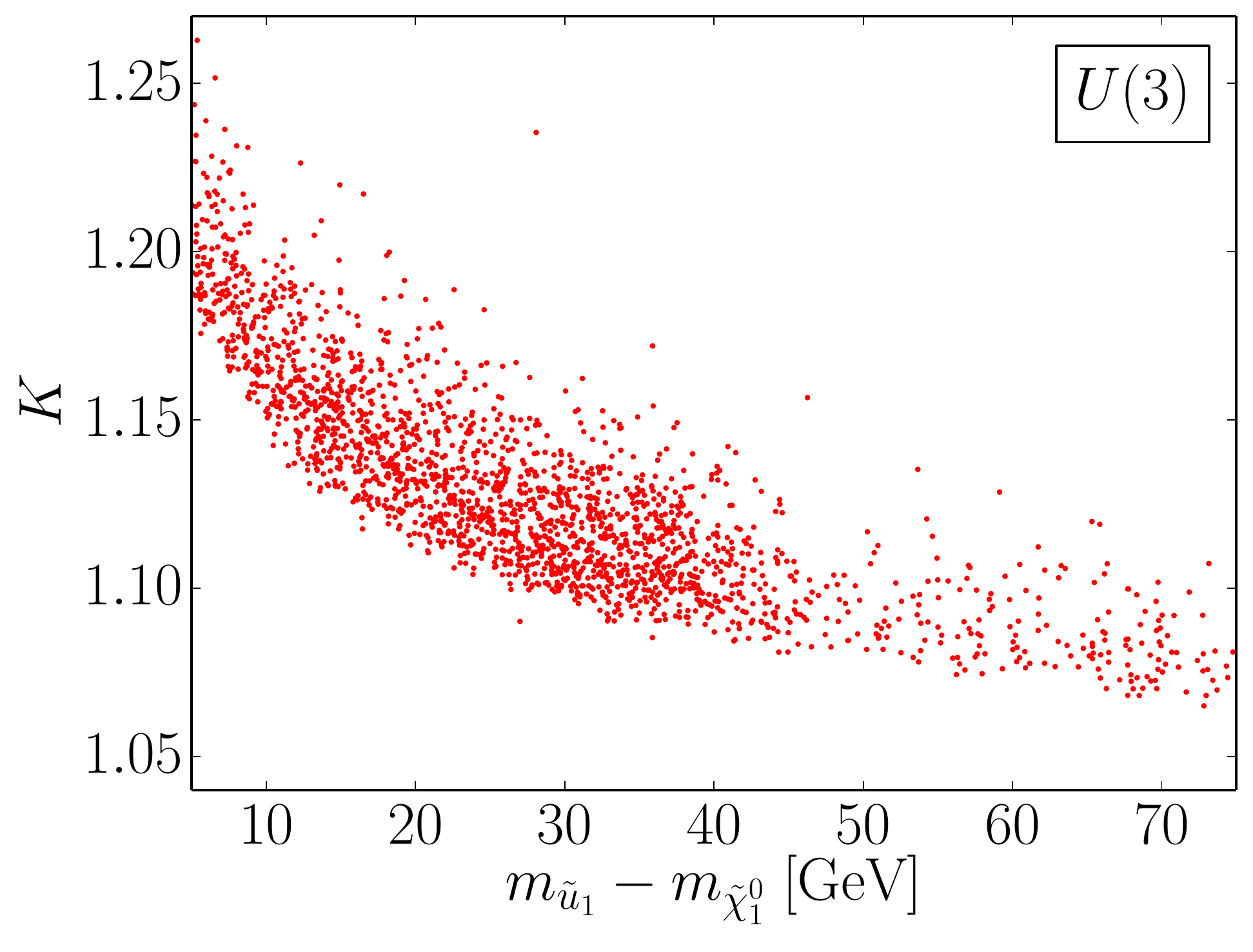}\hspace*{0.3cm}
  \includegraphics[width=7.7cm]{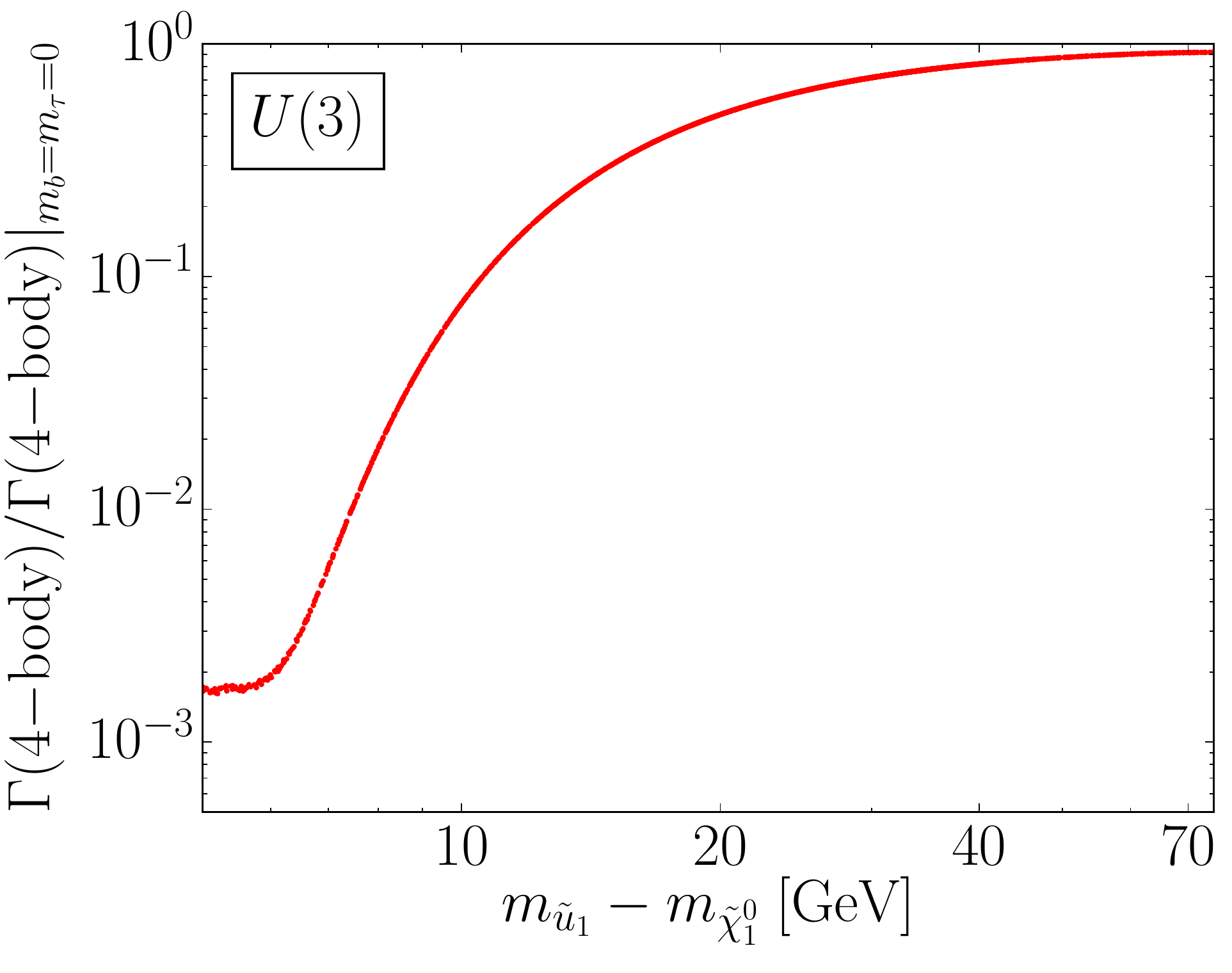}
 \caption{{\it Left:} $K$ factor for the decay $\tilde{u}_1\to c \tilde{\chi}_1^0$. 
 {\it Right:} The four-body decay width in full mass dependence $\Gamma(4$-body$)$ divided by the decay width where the third generation fermion masses are set to zero $\Gamma(4$-body$)|_{m_b=0, m_{\tau}=0}$. 
 \label{kfac}}
 \end{figure}

For flavour universal couplings at the Planck scale $\Lambda_{Planck}$ the dominant logarithmic terms $\log \Lambda_{Planck}/m_W$ with $m_W$ denoting the $W$ boson mass for the stop decay to $c\tilde{\chi}_1^0/u\tilde{\chi}_1^0$ have been first computed in \cite{Hikasa:1987db}. In \cite{Muhlleitner:2011ww} a full one-loop computation of the decay was done under the assumption of no FV at tree-level. This computation is hence restricted to the scale at which the couplings are flavour universal. In \cite{Grober:2014aha} (see also \cite{Aebischer:2014lfa}), the SUSY QCD corrections to the decay widths $\Gamma_{c\tilde{\chi}_1^0}$ and $\Gamma_{u\tilde{\chi}_1^0}$ allowing for a FV coupling at tree-level were computed. The result for the $K$ factor, defined as $K=\Gamma_{NLO}/\Gamma_{LO}$, is displayed in Fig.~\ref{kfac} (left panel).
 \subsection{Four-body decay $\tilde{u}_1\to d_i \tilde{\chi}_1^0 f \bar{f}'$}
 The four-body decays to $b\tilde{\chi}_1^0 f \bar{f}'$ have been computed for the first time in \cite{Boehm:1999tr}. In \cite{Grober:2014aha}  this computation was updated by including the mass dependence of the third generation fermions in the final states and by including FV, hence allowing in general for a final state $ d_i \tilde{\chi}_1^0 f \bar{f}'$, where $d_i$ denotes a down-type quark. The effect of including the mass dependence on the final state bottom quark and $\tau$ lepton is depicted in Fig.~\ref{kfac} (right panel). 
\subsection{Three-body decay $\tilde{u}_1\to d_i \tilde{\chi}_1^0 W$}
\begin{figure}
 \includegraphics[width=7.7cm]{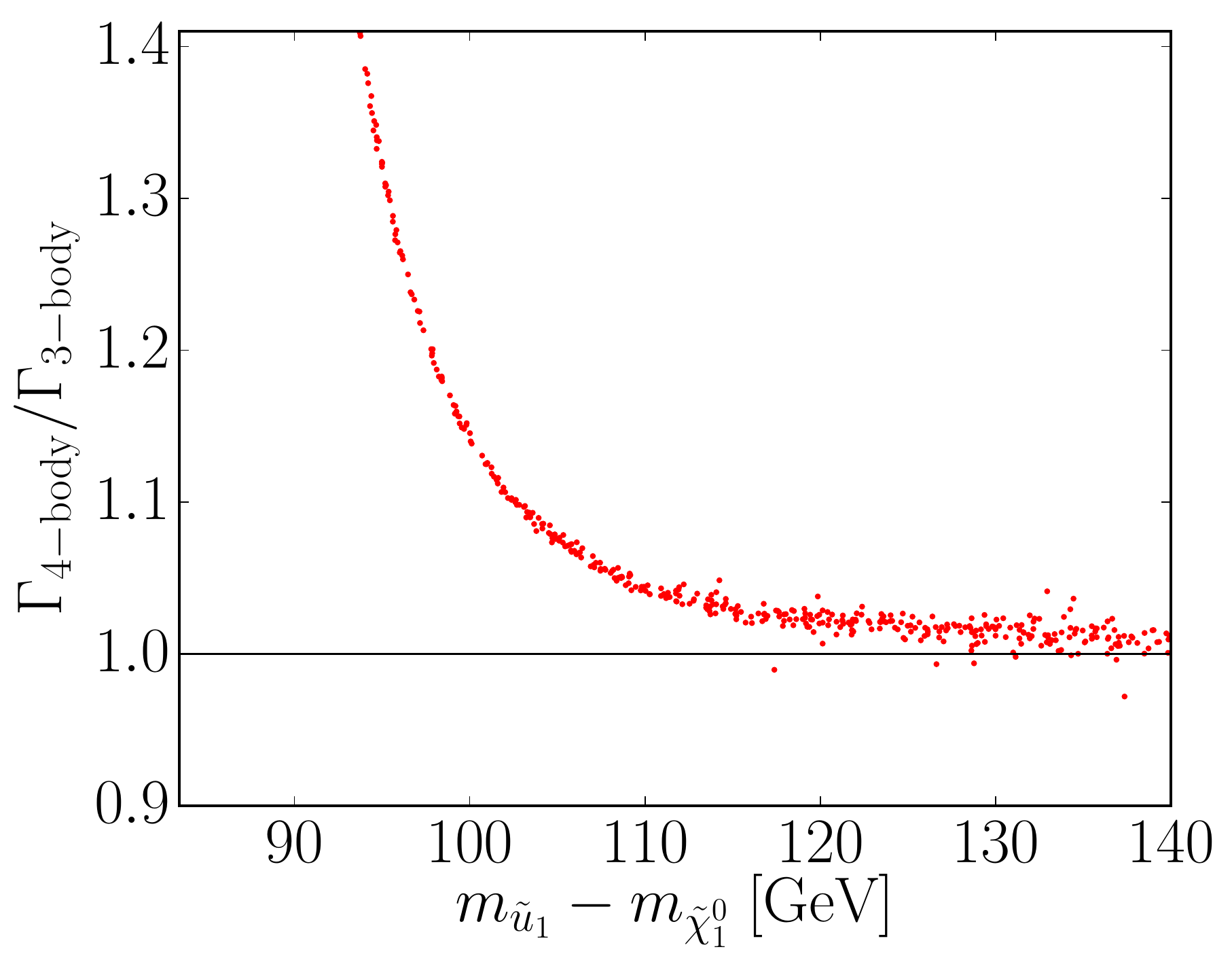}
  \includegraphics[width=8.15cm]{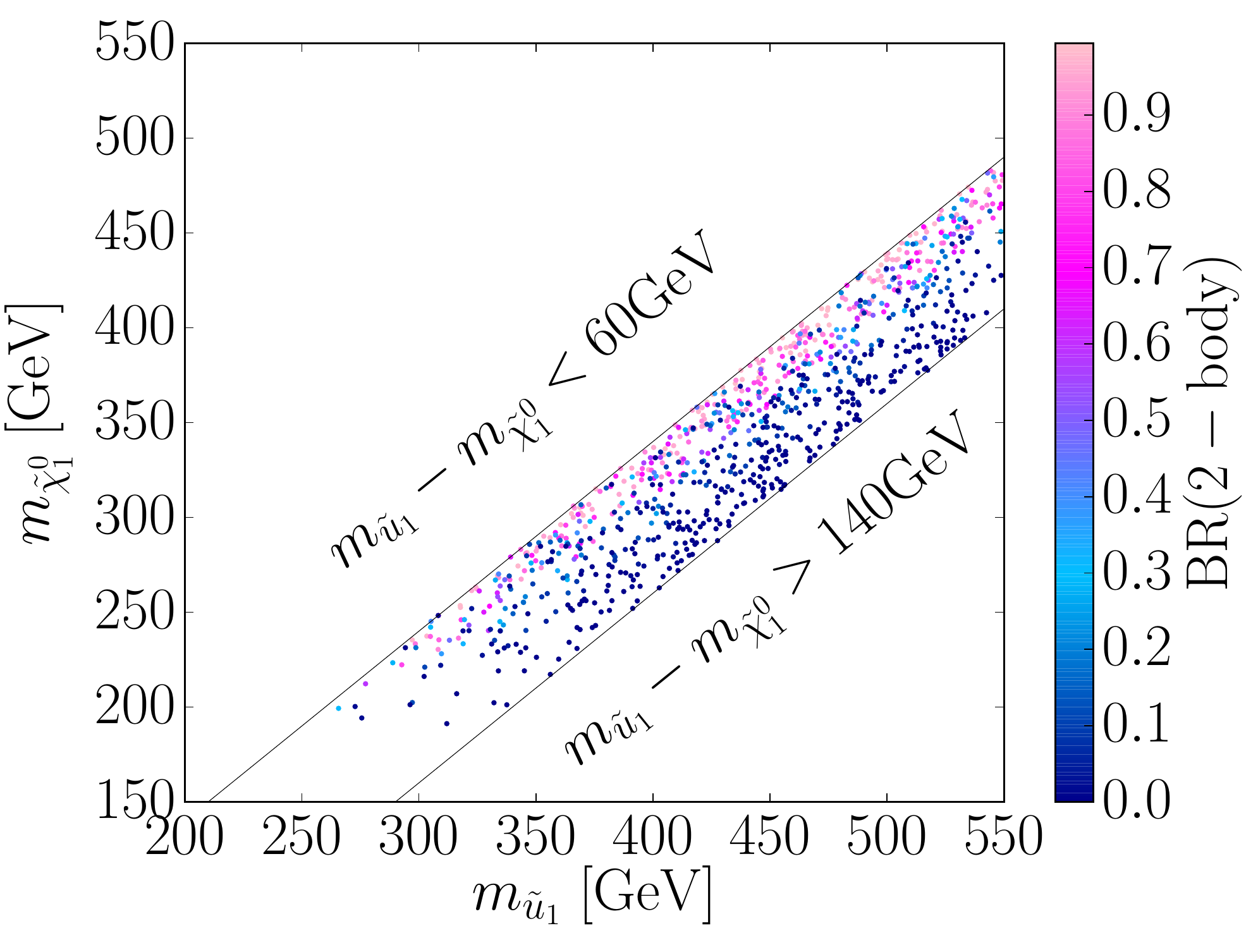}
 \caption{{\it Left:} Four-body decay width divided by three-body decay width for the stop decays to $d_i\tilde{\chi}_1^0 W^{(*)}$. {\it Right:} Scan over parameter space around the $W$ boson threshold. The color code indicates the size of the BR into the FV two-body decays. 
 \label{fouroverthree}}
 \end{figure}

If the mass difference between stop and neutralino $m_{\tilde{u}_1}-m_{\tilde{\chi}_1^0}$ becomes larger than the $W$ boson mass, the stop can decay via a three-body decay to  $d_i \tilde{\chi}_1^0 W$.
The three-body decay width to $b\tilde{\chi}_1^0W$ has been computed in \cite{Porod:1996at}. 
In \cite{Grober:2015fia} this was extended allowing for FV. In addition, off-shell effects were considered by incorporating a $W$ boson width into the four-body decay width of \cite{Grober:2014aha}. 
In Fig.~\ref{fouroverthree} the impact of these off-shell effects is shown. It can be inferred that for mass differences between stop and neutralino up to $m_W+30$ GeV the off-shell effects can be quite important. 
 \section{Numerical analysis of viable parameter space \label{numana}}
 In order to investigate the viable parameter space for light stops, a scan was performed. The SUSY spectrum has been generated by {\tt SPHENO} \cite{Porod:2003um}. Viability with Higgs data was checked by means of the codes {\tt HiggsSignals} \cite{Bechtle:2013xfa} and {\tt HiggsBounds} \cite{Bechtle:2008jh}, with the Higgs BRs and effective vertices obtained from {\tt HDECAY} \cite{Djouadi:1997yw}. With {\tt SuperIsoRelic} \cite{Arbey:2009gu} we checked  several flavour observables and that the relic density is not too large. The masses of the sparticles were chosen such that they escape the direct bounds of ATLAS and CMS. For $m_{\tilde{u}_1}-m_{\tilde{\chi}_1 ^0}<m_W$ we scaled down the exclusion limits \cite{Aad:2014kra} \cite{ATLAScharm} by the BRs and combined the different decay modes assuming that the BRs add up to one. Above the $W$ boson threshold we checked the exclusion limits for the decays to $b\tilde{\chi}_1^0W$ by means of the code {\tt SModelS} \cite{Kraml:2013mwa}. Limits for the decays into $c\tilde{\chi}_1^0$ above the $W$ threshold do not exist so far.
 \par
In Fig.~\ref{fouroverthree} (right panel) a scan for the $U(2)$ flavour assumption is shown for mass differences between stop and neutralino around the $W$ boson threshold. Apparently, even above the $W$ boson threshold the stop can still have a sizeable BR into $c\tilde{\chi}_1^0$. 
\par
\begin{figure}
 \includegraphics[width=8cm]{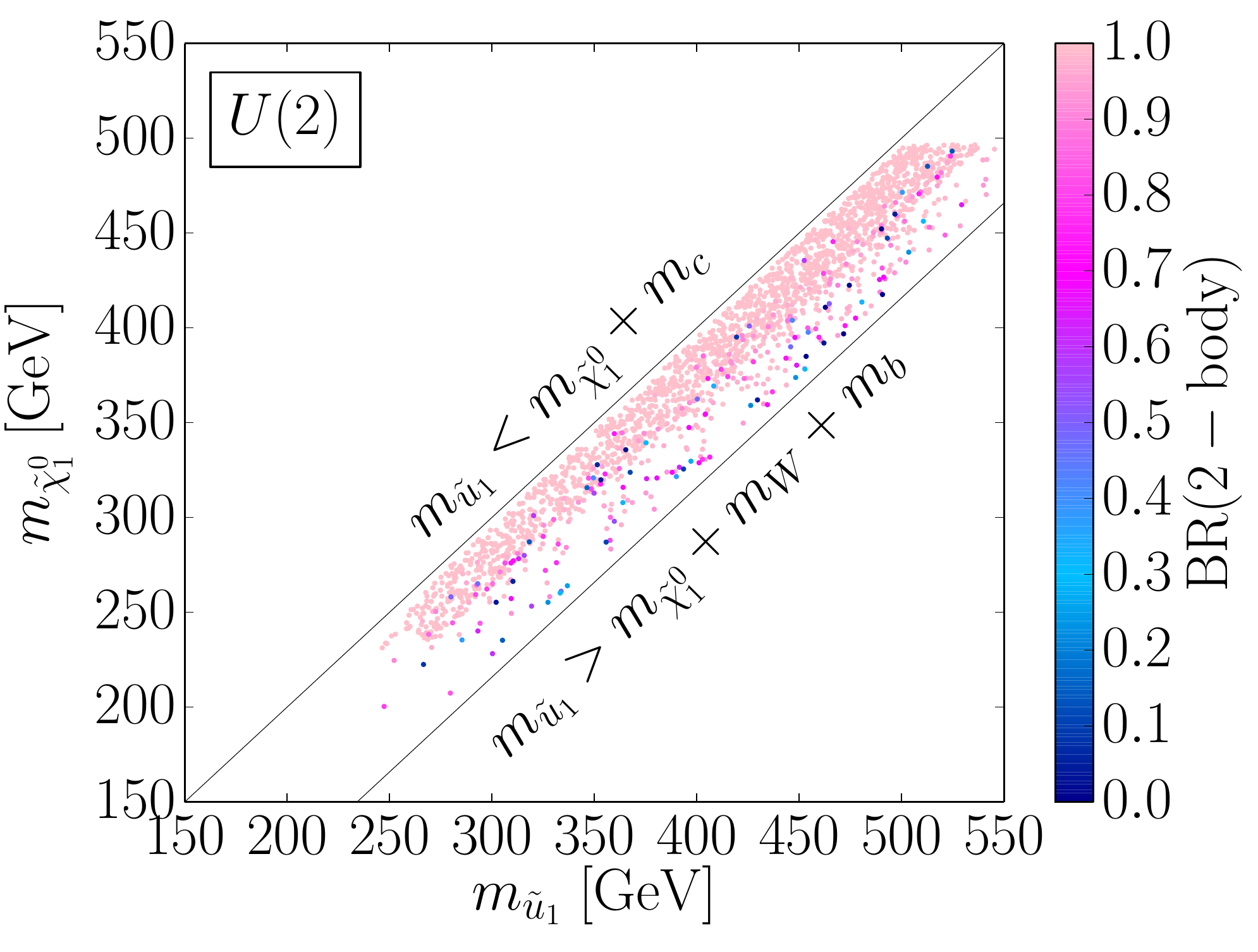}
  \includegraphics[width=8cm]{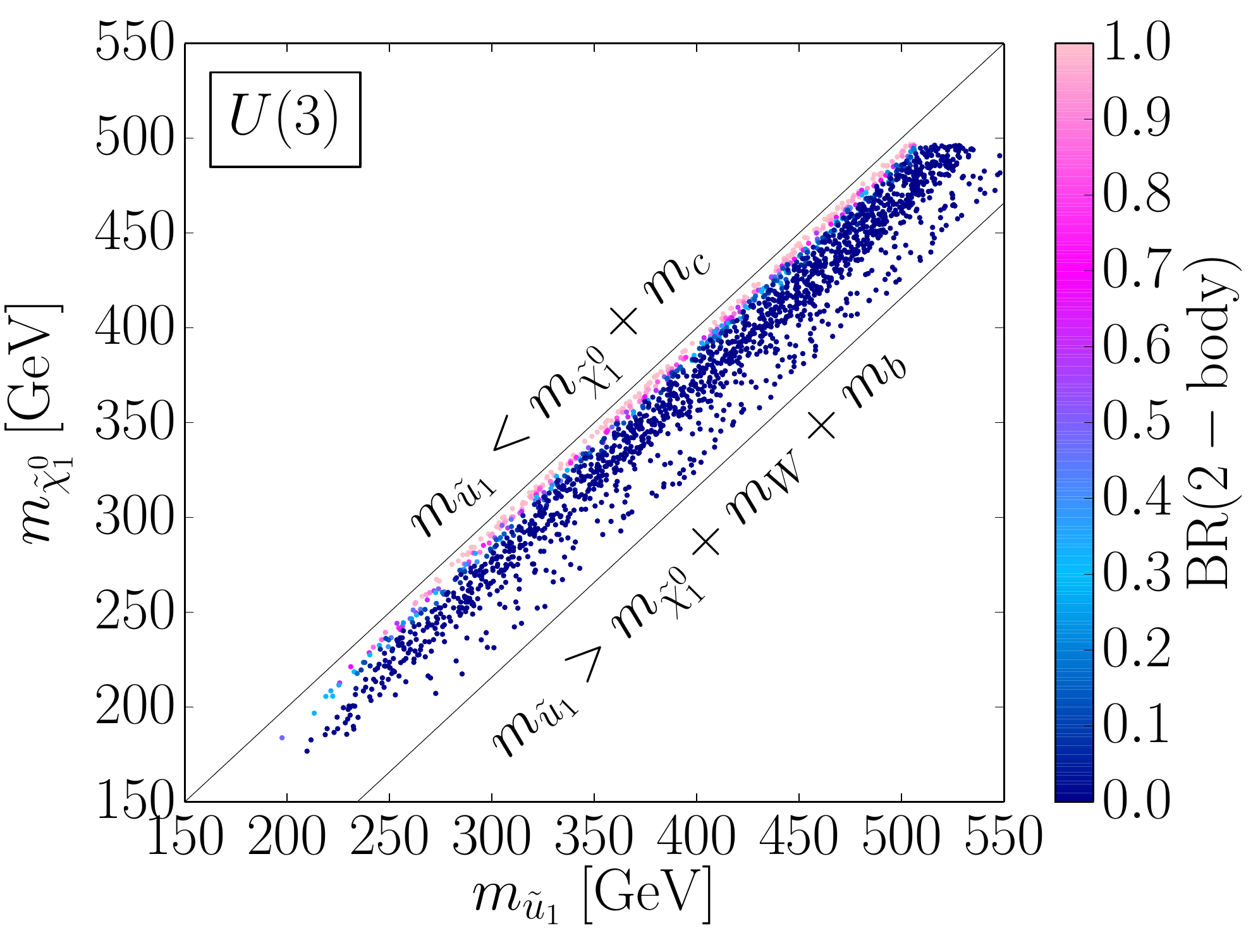}
 \caption{Scan over the parameter space for different flavour assumptions. The color code indicates the BR into the FV decay modes. 
 \label{inmassplane}}
 \end{figure}

 
In Fig.~\ref{inmassplane} a scan restricted to $m_{\tilde{u}_1}-m_{\tilde{\chi}_1^0}<m_W$ is shown. Two different flavour implementations are displayed, see sec. \ref{flavour}. If more FV is allowed the decay into $c\tilde{\chi}_1^0$ dominates, whereas for less FV the four-body decays dominate for mass differences larger than 15 GeV. There are also points where the stop has sizeable BRs in both decay channels, such that the experimental exclusion limits are reduced. Experimental exclusion bounds in terms of BRs are useful to compare to theory, a first attempt has been started in \cite{Aad:2015pfx}.

\section{Conclusion \label{conclusion}}
We have improved on the computation of the light stop decays by calculating SQCD corrections to the decays into $c\tilde{\chi}_1^0/u\tilde{\chi}_1^0$, inclusion of the mass dependence of third generation fermions in the four-body decays $d_i\tilde{\chi}_1^0f\bar{f}'$ and the off-shell effects for the three-body decays to $d_i \tilde{\chi}_1^0 W$. All of these decays were implemented into {\tt SUSYHIT} \cite{Djouadi:2006bz} allowing for general FV. A numerical analysis showed that BRs of the stop can significantly deviate from one, which lowers the experimental exclusion limits. For stop-neutralino mass differences larger than the $W$ boson mass, it should be considered in the experimental analyses that there can be sizeable BRs to $c\tilde{\chi}_1^0$. 
 \section{Acknowledgements}
 I am grateful to M.~M\"uhlleitner, E.~Popenda and A.Wlotzka for the fruitful collaboration and the organisers for a pleasant and stimulating atmosphere during the 27th Rencontres de Blois.

\end{document}